\documentclass[conference]{IEEEtran}
\IEEEoverridecommandlockouts
\pdfoutput=1
\usepackage{cite}
\makeatletter
\newcommand{\removelatexerror}{\let\@latex@error\@gobble}
\makeatother
\usepackage{hyperref}
\hypersetup{
    colorlinks=true,            
    linkcolor=blue,             
    filecolor=magenta,          
    urlcolor=cyan,              
    pdftitle={OpenDatasetAlbertaPowerGrid},
    pdfpagemode=FullScreen,
    }
\usepackage{amsmath,amssymb,amsfonts}
\usepackage{graphicx}
\usepackage{subfigure}
\usepackage{textcomp}
\usepackage{xcolor}

\usepackage{algorithmic}
\usepackage{bm}
\usepackage{diagbox}
\usepackage{booktabs}
\usepackage{url}
\usepackage{multirow}
\usepackage{geometry}
\newgeometry{
  top=72pt,
  bottom = 54pt,
  left = 54pt,
  right = 54pt
}

\usepackage{algorithm}
\usepackage{algorithmic}
\usepackage{array}
\usepackage{textcomp}
\usepackage{stfloats}
\usepackage{url}
\usepackage{diagbox}
\usepackage{verbatim}
\usepackage{graphicx}
\usepackage{multirow}
\usepackage{cite}
\usepackage{xcolor}
\usepackage{amsthm}

\usepackage{stfloats}

\usepackage{tikz}
\usetikzlibrary{positioning,arrows.meta} 
\usepackage{graphicx}

\def\BibTeX{{\rm B\kern-.05em{\sc i\kern-.025em b}\kern-.08em
    T\kern-.1667em\lower.7ex\hbox{E}\kern-.125emX}}

\usepackage{microtype}
\usepackage{float}
    
\begin{document}

\title{Open Datasets for Grid Modeling and Visualization: \\An Alberta Power Network Case}

\author{\IEEEauthorblockN{Ben Cheng}
\IEEEauthorblockA{\textit{Department of Computer Science} \\
\textit{University of Toronto}\\
Toronto, Ontario, Canada \\
bben.cheng@mail.utoronto.ca}
\and
\IEEEauthorblockN{Yize Chen}
\IEEEauthorblockA{\textit{Department of Electrical and Computer Engineering} \\
\textit{University of Alberta}\\
Edmonton, Alberta, Canada \\
yize.chen@ualberta.ca}}

\maketitle

\begin{abstract}
In the power and energy industry, multiple entities in grid operational logs are frequently recorded and updated. Thanks to recent advances in IT facilities and smart metering services, a variety of datasets such as system load, generation mix, and grid connection are often publicly available. While these resources are valuable in evaluating power grid's operational conditions and system resilience, the lack of fine-grained, accurate locational information constrain the usage of current data, which further hinders the development of smart grid and renewables integration. For instance, electricity end users are not aware of nodal generation mix or carbon emissions, while the general public have limited understanding about the effect of demand response or renewables integration if only the whole system's demands and generations are available. In this work, we focus on recovering power grid topology and line flow directions from open public dataset. Taking the Alberta grid as a working example, we start from mapping multi-modal power system datasets to the grid topology integrated with geographical information. By designing a novel optimization-based scheme to recover line flow directions, we are able to analyze and visualize the interactions between generations and demand vectors in an efficient manner. Proposed research is fully open-sourced and highly generalizable~\footnote{\url{https://github.com/BenCheng2/CarbonDistributionMap}}, which can help model and visualize grid information, create synthetic dataset, and facilitate analytics and decision-making framework for clean energy transition.
\end{abstract}

\begin{IEEEkeywords}
Correlation Modeling, Graph theory, Linear Programming, Power flow, Topology,  Transmission modeling,Visualization
\end{IEEEkeywords}

\section{Introduction}

The evolution of smart grid and sustainable power systems call for fine-grained modeling and visualization of power grid data including thousands of nodes and timesteps. This necessity is particularly critical given the increasing complexity of power generation and consumption patterns, along with the growing need for more advanced programs such as electricity market design, demand response, geographical load shifting, inverter-based resources integration, and carbon emission assessments~\cite{iea2023}. 

\begin{figure}
    \centering
    \includegraphics[width=1\linewidth]{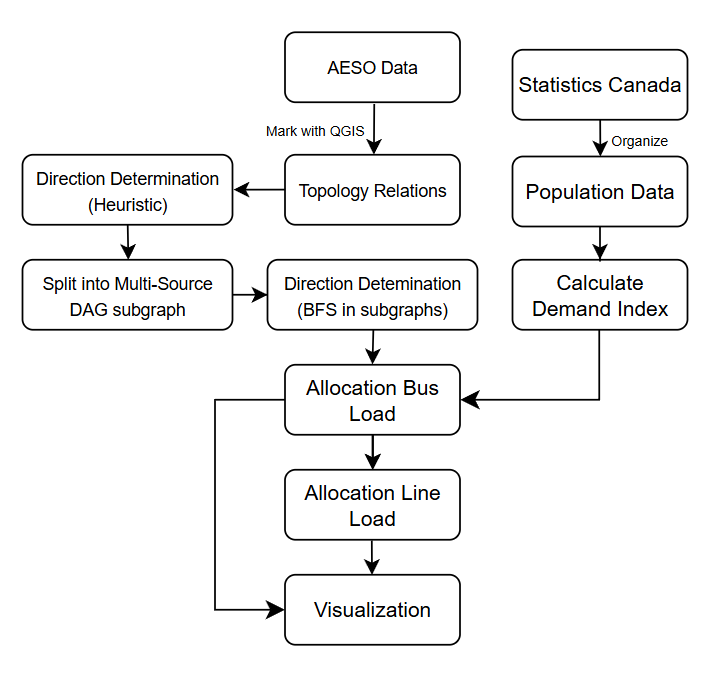}
    \caption{Procedure overview. By only using publicly available datasets and API, we are able to recover transmission network topology, determine lineflow directions and visualize geographical characteristics.\vspace{-10pt}}
    \label{fig:flowchart}
\end{figure}

In today's power system practice, grid modeling and visualization are usually implemented in a separate manner by each region's system operator. Some North American and European regional transmission operators (RTOs) either publish zonal demand (ISO New England and New York ISO) or discrete pricing points (California ISO).  Contour maps are also provided by RTOs such as Midwest ISO, PJM, and Southwest Power Pool. These information are published either for real-time or real-time and day-ahead (New York ISO, ISO Enw England, and ENTSO-E). Recent advancements in power system data visualization also help understand grid dynamics and comprehensive interactions between various components and load regions~\cite{overbye2019wide, mao2019w4ips}.

However,  to analyze the power grid's ever-changing patterns of generation, demand, and carbon emissions, engineers and electricity end users often lack effective and efficient data analytical tools~\cite{cheng2023gridviz}. One major obstacle is that many of power grid data are not readily available either due to infrastructure limitations or privacy concerns. This lack of accessible data can hinder the accurate modeling of power system dispatch, renewables integration, and emissions associated with power flow, leading to discrepancies between theoretical analysis and actual power system operations.  For instance, grid connection is often publicized as an image rather than detailed topology, and power generation and demand are only available for the entire or regional system, which is actually comprised of hundreds of nodes and subregions. Previous work often look into each independent subproblem in topology identification~\cite{li2013blind, medjroubi2017open}, line parameter estimation~\cite{knyazkin2004parameter}, demand modeling~\cite{arjmand2022canada} and etc. Some research use statistical learning on fine-grained, high-resolution power measurements to infer the topology~\cite{deka2023learning}. While in \cite{taylor2023california}, the topology and line parameters are recovered for the California grid based on modeling market clearing process through optimal power flow. Researchers also utilize OpenStreetMap and a least-cost routing algorithm to trace   powerlines in distribution network~\cite{arderne2020predictive}, while it is challenging to infer the directed network topology by taking demand patterns into account. 

In this paper, we ask the following research questions:

\emph{How to use publicly available open data to model power networks? What are the barriers in achieving realistic modeling and visualization?}

To answer these questions, we design a data collection and analytics pipeline and use Alberta, Canada grid as a working example.  By identifying a combination of multimodal datasets from system operators, statistical reports and maps, it allows us to approximate and simulate the critical dynamics of power generation, transmission, demand, and supply throughout the geographical region. Though such open data is not perfect in terms of preciseness and resolution, our designed approach is generalizable, and is able to trace the generation and demand pattern in a ``practical but not exact'' manner, as private information on generators and line parameters are not exposed. The overall procedure is illustrated in Fig. \ref{fig:flowchart}. In addition, we also find that quality of electricity system modeling heavily relies on input data, which calls for more transparent data sharing and open-sourcing grid modeling practices.

\section{Problem Formulation}
In this work, we denote the power network as a graph of $\mathbf{G}=(\mathbf{B},\mathbf{E})$, where $\mathbf{B} \in \mathbb{R}^n$ denotes the vector of $n$ buses, $\mathbf{E}\in \mathbb{R}^m$ denotes the vector of $m$ powerlines. In addition, let $b_i^g$ and $b_i^l$  denote the generation and load bus respectively. $v_{i}$ denotes the voltage level at bus $i$, and $f_{ij}$ denotes the line flow value. We focus on transmission grid in this work,  and the task of recovering topology and line flow directions using public data faces several challenges due to data availability:
\begin{itemize}
    \item \textbf{Insufficient Demand and Supply Data:} Ideally, complete data on the demand and supply at nodal level would allow for accurate modeling of the network's power distribution. However, such information is often proprietary and not publicly available;
    \item \textbf{Unknown Line Flow Directions:} The direction of power flows, critical for modeling the supply from generators to each node, frequently changes and is also not publicly accessible. This lack of directional data complicates the ability to model the actual flow paths and operational dynamics of the network;
    \item \textbf{Standalone Data Files:} The network topology, generator metadata, and substation metadata and load data are all standslone and separated, making it challenging to analyze grid information in a cohesive approach.
    \end{itemize}


To address these challenges, we adopt several strategic measures to ensure our method's effectiveness. This approach enables us to circumvent the constraints posed by national security and data variability across different facilities, facilitating a more robust analysis despite data limitations.

\section{Description of Datasets}
In this Section we describe the data sources we identify, which are all publicly available and can be queried in real-time. A summary of these datasets are shown in Table \ref{tab:datasets}.
\newcolumntype{P}[1]{>{\centering\arraybackslash}p{#1}}
\begin{table*}[h]
    \centering
    \renewcommand{\arraystretch}{1.2} 
    \begin{tabular}{P{3cm}|P{7cm}|P{2.5cm}|P{3cm}}
        \hline
        \textbf{Dataset Name} & \textbf{Dataset Description} & \textbf{Dataset Source} & \textbf{Processed Variables} \\
        \hline
        Alberta Interconnected Electric System Map & The AIES‑Map is a publicly available raster map that depicts the approximate locations of major high‑voltage substations, transmission lines, municipal boundaries, and planning area delineations across Alberta. & AESO Assets \cite{aiesmap} & $G=(\mathbf{B,E})$; Geolocations for $b_i^g$, $b_i^l$ \\
        \hline
        Single Line Diagram (SLD) &  The SLD is a comprehensive schematic that depicts all major substations and transmission lines across Alberta, providing a province‑wide overview of the existing grid infrastructure. & AESO Assets \cite{singlelinediagram} & $G=(\mathbf{B,E})$; Voltage level $\mathcal{V}^\mathcal{B}$, $\mathcal{V}^\mathcal{E}$\\
        \hline
        Hourly load by area  & Hourly load by AESO planning area for Jan-2011 through Dec-2024 & AESO Market and system reporting \cite{albertahourload}. & $\mathcal{H}:=\{\mathbf{p}:h_{\mathbf{p}} \}$ \\
        \hline
        Current Supply Demand Report & The current generation and maximum capacity of each generator within Alberta & AESO Market Report \cite{currentdemandsupply} & $\mathcal{G}:=\{b_i: g_{b_i}\}$ \\
        \hline
        Population and dwelling counts & The population of each city in Alberta in 2021 & Statistics Canada \cite{populationcensus} & Population in each planning area \\
        \hline
        
    \end{tabular}
    \caption{Summary of publicly-availabe datasets used and processed in this study.}
    \label{tab:datasets}
\end{table*}

\subsection{AESO Data}
\begin{figure}[H]
    \centering
    \includegraphics[width=1\linewidth]{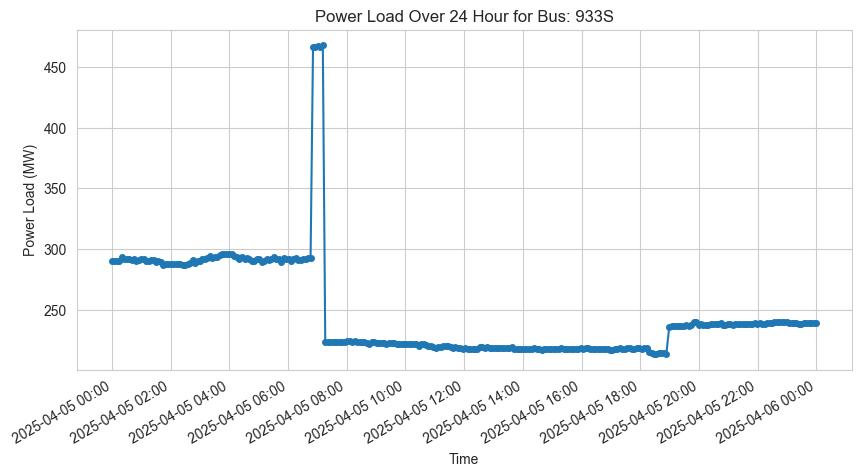}
    \caption{Bus Load 933S in 24 hours, which is recorded by continuously monitoring AESO data repository for 24 hours (2025.04.05).}
    \label{fig:trend-24hour}
\end{figure}

The Alberta Electric System Operator (AESO) organizes its transmission management and planning tasks within numbered planning areas across Alberta. Each area (e.g. Area 4 = Medicine Hat, Area 60 = Edmonton) encapsulates the local characteristics of the power system, facilitating the identification and assessment of necessary transmission upgrades. Additionally, these areas are aggregated into larger planning regions (e.g. Northwest, Northeast) which aid in conducting province-wide system studies. However, public data provided by AESO has several caveats. For instance, in the \textbf{AIES‑Map \cite{aiesmap}}, AESO and the provincial government do not provide precise geospatial coordinates for individual substations, generation facilities, or line alignments. To tackle that, spatial data used in this study are digitized directly from this map. In addition, nodal load data are only available in real time without historical records (Fig. \ref{fig:trend-24hour}), while only historical system-level demand are archived at AESO data repository.

We import the AIES-Map raster into QGIS and map each substation as a point and each transmission corridor as a polyline into separate vector layers. In cases where facilities are absent or incomplete on the \textbf{AIES-Map}, we refer the \textbf{Single Line Diagram (SLD)} to approximate their locations. All digitized features are then exported in the EPSG:3857 (Web Mercator) 2D projection. 

Due to the unavailability of the underlying coordinate reference system, we assume EPSG:26911 is adopted~\footnote{\url{https://epsg.io/26911}}, and our spatial data rely on two-dimensional pixel positions instead of geographically calibrated coordinates. This method may introduce slight geometric distortions in our visualizations, while we find that such a procedure does not affect the overall spatial trend in Alberta's power network and the precision of power dispatch calculation.



\begin{figure*}
    \centering
    \includegraphics[width=0.99\linewidth]{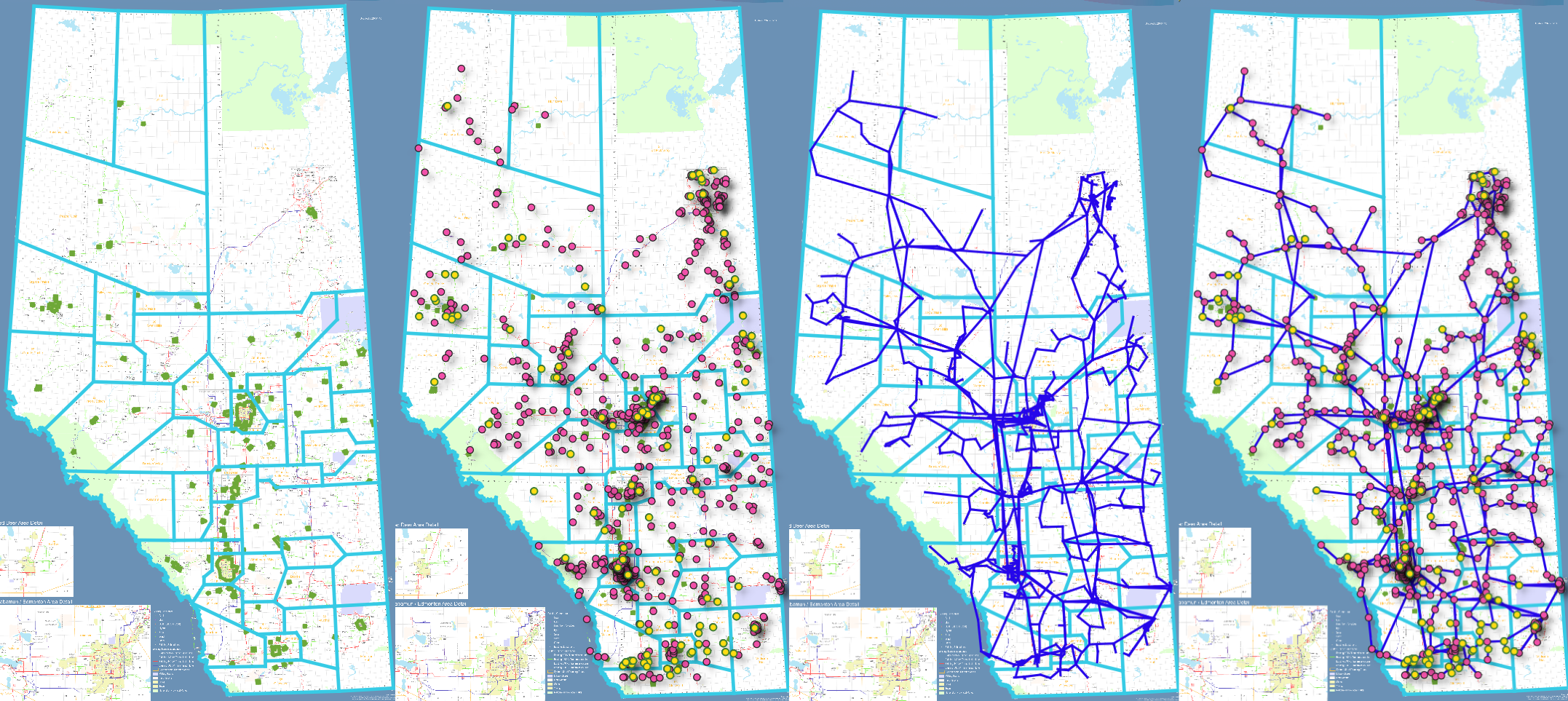}
    \caption{Proposed visualization for (a) City (\textcolor{green}{green}) and Planning Area (\textcolor{cyan}{cyan}) Border (b) Bus (\textcolor{red}{red}) and Generator (\textcolor{yellow}{yellow}) Location (c) Powerlines (\textcolor{blue}{darkblue}) Infrastructure and Connection (d) Rich Power Grid Information.}
    

    \label{fig:qgis-visual}
\end{figure*}

\subsection{QGIS Data}
QGIS is an open-source Geographic Information System (GIS) software, allowing users to analyze and visualize spatial data. It stands out from ArcGIS, its commercial counterpart, by being fully open-source, which supports extensive customization and community-driven enhancements. We utilized this software to mark and preprocess the spatial information for buses and powerlines within Alberta.

The spatial infrastructure datasets (Substation.csv, Line.csv, CityBorder.csv, PlanningAreaBorder.csv, and CityPopulationPoint.csv) are exported directly from the vector layers of our QGIS project, representing all digitized substations, transmission corridors, municipal boundaries, AESO planning area boundaries,s and urban boundaries within Alberta (Fig.\ref{fig:qgis-visual}.a). The publicly available generator and substation dataset corresponds to the naming and location of generators derived from SLD, and also includes Maximum Capacity and the Generator Type (Fig.\ref{fig:qgis-visual}.b). For the transmission line data, it corresponds to the line connection derived from SLD, while also providing each line's voltage level (Fig.\ref{fig:qgis-visual}.c). We also use city-level population data, which contain population and area information for Alberta municipalities only, extracted from the "Statistics Canada's 2021 Census: Population and dwelling counts" dataset. Once all datasets have been processed, we also use QGIS to stack the layer of city border, generation and demand nodes, and power networks in one map, which can effectively visualize rich power grid information (Fig.\ref{fig:qgis-visual}.d).


\subsection{Demand Index}
Once the AESO and QGIS data are processed, we now look into nodal-level demand modeling and line-level power dispatch modeling. Due to the lack of precise bus-level load measurements $b_i^l$, we devise Demand Index $\mathcal{I}$ to disaggregate planning-area demand to the individual bus.  The data utilized for this estimation comes from the the average hourly load for each AESO planning area over the period from 2011 to 2024. We follow the proportional share heuristic to allocate demand based on node location and regional population. 

We hypothesize and demonstrate a strong correlation between population distribution and power demand across various planning areas. For the hypothesis testing, we aggregate city-level population data to determine the total population for each planning area $pa$, we then organize these populations as a proportion in a ratio vector for further comparison:
\begin{align*}
\mathbf{v} = \left( \text{population(pa}_1), \text{population(pa}_2), \dots, \text{population(pa}_n) \right);
\end{align*}

We then organize the hourly power load ratios into a vector, which serves as proxies for the power demand distribution:
\begin{align*}
\mathbf{v}^{'} = \left( h_{\text{pa}_1}, h_{\text{pa}_2}, \dots, h_{\text{pa}_n} \right).
\end{align*}

For our analysis, we compare \(\mathbf{v}_{\text{year}}'\) (hourly load ratio vector for specific years between 2012 and 2024) with \(\mathbf{v}\) (2021 population distribution ratio), using both cosine similarity and the Pearson Correlation Coefficient in Table \ref{tab:similarity}. Our findings reveal that \(\mathbf{v}_{\text{2021}}'\) (2021 hourly load ratio vector) exhibit the strongest correlation with \(\mathbf{v}\). Specifically, the population proportion vector demonstrate a cosine similarity of 0.8959 and a Pearson Correlation Coefficient of 0.9080 with the 2021 average hourly power consumption, both nearing 90\%. These exceptionally high metrics confirm a very strong similarity between \(\mathbf{v}\) and \(\mathbf{v}_{\text{2021}}'\), supporting the validity of utilizing population distribution to model power demand distribution.


\begin{table}[htbp]
\centering
\begin{tabular}{c |c |c}
\toprule
Year & Cosine Similarity & Pearson Correlation Coefficient \\
\midrule
2012 & 0.7722 & 0.7579 \\
2013 & 0.7583 & 0.7396 \\
2014 & 0.7465 & 0.7242 \\
2015 & 0.7280 & 0.7005 \\
2016 & 0.7311 & 0.7045 \\
2017 & 0.7026 & 0.6682 \\
2018 & 0.6719 & 0.6301 \\
2019 & 0.6624 & 0.6183 \\
2020 & 0.8100 & 0.8025 \\
2021 & 0.8959 & 0.9080 \\
2022 & 0.8944 & 0.9058 \\
2023 & 0.8956 & 0.9054 \\
2024 & 0.8908 & 0.9002 \\
\bottomrule
\end{tabular}
\caption{Similarity between Hourly Power Load Ratio and Population Ratio from 2012 to 2024.}
\label{tab:similarity}
\end{table}

\section{Tree-Based Directed Network Recovery}
\label{sec:direction}
As line parameters and generator parameters can be highly sensitive and private information, in this work we do not solve exact power flow problems. Alternatively,  we propose to recover the line flow directions and transmitted power throughout the whole network. Based on the network topology and the physical characteristics of the network infrastructure, we develop a set of heuristic rules to determine the most likely direction of power flows along each line.  Although power flows can change in real time based on immediate demand and supply, we find our approach can accommodate all generations and a number of load variations. To be specific, the following modeling assumptions are taken into account:
\begin{itemize}
\item \emph{Two-end-Voltage Heuristic}: Power should flow from the bus with high voltage to low voltage;
\item \emph{Line-Voltage Heuristic}: Power should not flow from a bus with low voltage into a power line with high voltage;
\item \emph{Generator-as-a-Source Heuristic}: For any power line connecting a bus with a generator, power should flow out of the generator. (If $p_i^g=0$, this heuristic does not apply).
\end{itemize}

We also note that within Alberta’s transmission network, 500kV circuits serve predominantly long‑distance transfer rather than transmitting power directly from the generators - there are only minor generators connected with 500kV-level lines. In contrast, the 138kV and 240kV networks function as regional transmission backbones that more closely aggregate generation and load. For modeling purposes, it is reasonable to process power flows between 240kV and 500kV buses and lines as effectively interchangeable, reflecting the role of 500kV infrastructure in interregional transmission.

There are 4 instances where Power Line Voltage $<$ Substation Voltage on both sides. 
One of these (500kV → 240kV) is resolved by the interchangeable rule described before, and the remaining 3 cases represent atypical conditions: either direct transfers of power from high‑voltage transmission lines into the regional network or arising from misrecording of data or omitted intermediary infrastructure that cannot be accurately depicted on connection maps. Since there are only limited occurrences of these special cases, for modeling purposes, we assume power can flow freely between terminal substations.

In addition, there are two cases  where two connected substations both have generators. The presence/absence of the transmission line between them does not alter overall network connectivity or supply for other substations, it is treated as a random direction to adapt our model.

Based on heuristic rules, we can determine the directions for a significant portion of power lines (355 out of 855). The remaining transmission lines, with undetermined directions, form multiple subgraphs. Each subgraph may have one or more entry points, where higher voltage or generated power flows into this Multi-Source Directed Acyclic Graph (DAG). Subsequently, by initiating a breadth-first search (BFS) from these entry points, we can determine the direction of every line within the DAG by iterating over them through the BFS algorithm. Since BFS primarily targets nodes, the directions of some power lines may be skipped; these are added back after BFS completes with random assignment, as they do not impact the reachability of the buses. For each subgraph, power flows from the inflow nodes to all other nodes, including the outflow nodes, ensuring that the inflow and outflow of each subgraph align with the external network. This approach ensures that the established directions are consistent and do not conflict with those outside the subgraph. See details of line direction determination in Algorithm \ref{alg:line_dir}.

 \begin{algorithm}
 \caption{Direction Decision Algorithm: Strongly Connected Component}
 \begin{algorithmic}[1]
 \renewcommand{\algorithmicrequire}{\textbf{Input:}}
 \renewcommand{\algorithmicensure}{\textbf{Output:}}
 \REQUIRE $\mathcal{V}^\mathcal{B}$:= Buses Voltage, $\mathcal{V}^\mathcal{E}$:= Powerlines Voltage Level
 \ENSURE $\mathcal{D}$:= Powerlines Direction

    \FOR {$e$ in $\mathcal{E}$}
        \STATE Determine $d_e$ by Three Heuristics in Sec. \ref{sec:direction} 
        \IF {$d_e$}
          \STATE $\mathcal{D}$.add($d_e$)
        \ENDIF
    \ENDFOR

    \STATE Subgraphs $\gets$ GetSubgraph($\mathcal{E} - \mathcal{D}$)

    \FOR {subgraph in Subgraphs}
        \STATE $\mathcal{B}^{input} \gets$ bus with the highest voltage or generators
        \STATE $\mathcal{Q} \gets$ Queue(); $visited \gets$ Set()
        \FOR {$b$ in $\mathcal{B}^{input}$}
            \STATE $\mathcal{Q}$.enqueue($b$); $visited$.add($b$)
        \ENDFOR
        \WHILE {$\mathcal{Q}$}
            \STATE $u \gets \mathcal{Q}$.dequeue()
            \FOR {$v$ in neighbor($u$)}
                \IF {$v$ in $visited$}
                    \STATE $\mathcal{Q}$.enqueue($v$); $visited$.add($v$); \STATE $\mathcal{D}$.append($d_{(u,v)}$)
                \ENDIF
            \ENDFOR
        \ENDWHILE
    
    \ENDFOR

  \RETURN $\mathcal{D}$
 \end{algorithmic} 
  \label{alg:line_dir}
 \end{algorithm}

\section{Power Dispatch Modeling and  Visualization}
Based on the recovered topology of Alberta power grid and nodal demand, we are able to analyze the power dispatch patterns and conduct model-based visualizations.
\subsection{Nodal Allocation of Power Demand}
According to Statistics Canada's 2021 Census: Population and dwelling counts \cite{populationcensus}, 84.8\% of Alberta's residents reside in urban areas. Population and electricity demand demonstrate strong statistical alignment (cosine similarity = 0.895; Pearson’s r = 0.908), justifying the strategy of assigning bus demand based on population distribution. Consequently, we could assign 84.8\% of the estimated demand to urban buses.


 Based on the municipal boundaries defined in the AIES Map (see Fig~\ref{fig:qgis-visual}), we delineate urban and non-urban areas within Alberta to estimate a fine-grained demand allocation at each bus. 
 To distribute this demand, we first calculate the total demand for each planning area based on hourly load data, which serves as the RDI (Relative Demand Index). This approach allows us to abstract from the actual magnitude of the load, focusing instead on its distribution across the urban and non-urban areas within the planning area. For each planning area, the total urban RDI is obtained by multiplying the area's total RDI by 84.8\%, and similarly, the non-urban RDI by 15.2\%. This demand is then evenly distributed among the buses falling into the urban and non-urban areas, respectively. For instance, considering Alberta's total load and population, if a planning area has a RDI of 100 (generated from 100 MW hourly power load), 84.8 unit of RDI would be allocated to urban substations and 15.2 unit of RDI to non-urban substations. The allocation per bus within these categories would then depend on the number of buses in each category.

After the city/non-city allocation, we average the RDI to each substation based on geographical regions. For instance, consider a planning area with an RDI of 100, which represents the combined total demand for urban and non-urban areas. If this area contains 5 urban buses and 3 non-urban buses, then urban buses collectively receive $100 \times 84.8\%$ = 84.8 units of the total RDI. Under uniform distribution, each urban bus would therefore receive $\frac{84.8}{5} = 16.96$ units of RDI. Similarly, the non-urban buses collectively receive $100 \times 15.2\%$ = 15.2 units of the total RDI. Each non-urban bus would then receive $\frac{15.2}{3} \approx 5.07$ units of RDI. See Algorithm \ref{alg:demand} for details.

\begin{algorithm}
\caption{Estimated Demand Allocation}
\begin{algorithmic}[1]
\renewcommand{\algorithmicrequire}{\textbf{Input:}}
\renewcommand{\algorithmicensure}{\textbf{Output:}}
\REQUIRE $\mathcal{P}$:= set of $planning\_area$, $\mathcal{H}:=\{\mathbf{p}:h_{\mathbf{p}} \}$ for the dictionary of planning area and average hourly load. 
\ENSURE $\mathcal{I}$: Demand Index for bus

\STATE $\mathcal{I}$ = dict()

\FOR {$\mathbf{p}$ in planning\_areas}
    \STATE $\mathcal{D}_{u} \gets \pi_{\text{urban}} \times h_{\mathbf{p}}$
    \STATE $\mathcal{D}_{nu} \gets (1 - \pi_{\text{urban}}) \times h_{\mathbf{p}}$
    \STATE $\mathcal{B}^{\mathbf{p}} \gets $ Bus in area $\mathbf{p}$
    \STATE $\mathcal{B}_{u}^{\mathbf{p}} \gets \mathcal{B}^{\mathbf{p}}$ in urban area
    \STATE $\mathcal{B}_{nu}^{\mathbf{p}} \gets \mathcal{B}^{\mathbf{p}}$ not in urban area
        
    \FOR {$b$ in $\mathcal{B}^{\mathbf{p}}$}
        \IF {$b$ in urban area}
            \STATE $\mathcal{I}[b]$ = $\mathcal{D}_{u}$ / $|\mathcal{B}_{u}^{\mathbf{p}}|$
        \ELSE
            \STATE $\mathcal{I}[b]$ = $\mathcal{D}_{u}$ / $|\mathcal{B}_{nu}^{\mathbf{p}}|$
        \ENDIF
    \ENDFOR
\ENDFOR
\RETURN $\mathcal{I}$
\end{algorithmic} 
\label{alg:demand}
\end{algorithm}


\begin{algorithm}
\caption{Estimated Substation Power Load}
\begin{algorithmic}[1]
\renewcommand{\algorithmicrequire}{\textbf{Input:}}
\renewcommand{\algorithmicensure}{\textbf{Output:}}
\REQUIRE $\mathcal{I}$:= Demand Index for buses, $\mathcal{B}^g$:= Generation buses, $\mathcal{G}:=\{b_i: g_{b_i}\}$ for the dictionary of buses and generation. 
\ENSURE $\mathcal{L}$: Power load for buses

\STATE $\mathcal{L} \gets$ dict()

\FOR {$b$ in $\mathcal{B}^g$}
    \STATE $\mathcal{B}^r$ = get\_reachable\_station(g)
    
    \FOR {$b_{r}$ in $\mathcal{B}^r$}
        \STATE $p$ = $\mathcal{I}[b_{r}]$ / sum([$\mathcal{I}[b_{r'}]$ for $b_{r'}$ in $\mathcal{B}^r$])

        \STATE $\mathcal{L}[b_r]$ += $p$ $\times$ $g_{b}$
    \ENDFOR

\ENDFOR
    
\RETURN $\mathcal{L}$
\end{algorithmic} 
\end{algorithm}

After estimating the power demand for each substation within the designated planning areas, we are able to find how such power is supplied by generators throughout the network. To achieve such a goal, we develop a Linear Programming (LP) model to calculate the power flows on each transmission line with known line flow directions. The goal of this model is to determine the load on each powerline starting from the load at each substation, ensuring that the power distribution aligns with our initial determination of topology and line directions. For each bus $b$ in $\mathcal{B}$, we impose a flow‑conservation constraint ensuring that total inflow and nodal generation equals to sum of total outflow and nodal consumption. Mathematically, we want to find power dispatch and line flow values so as to minimize the summation of nodal mismatch:

\begin{subequations}
\begin{align}
    \hspace{-80pt} \min_{b_i^g, f_{ij}} \quad &\sum_{i \in \mathcal{B}} \epsilon^+_i\\
    s.t. \quad & \sum_{j : (j,i) \in L} f_{j,i} + b_i^g - \sum_{j : (i,j) \in L} f_{i,j} - b_i^l = \epsilon^+_i \label{equ:balance}\\
& \epsilon^+_i \geq 0 \quad \forall \ i \in \mathcal{B}   \label{equ:non-negative} \\
&f_{i,j} \geq 0 \quad \forall \ (i,j) \in \mathcal{E} \label{equ:lineflow}
\end{align}
\end{subequations}
where \( \epsilon^+_i \) denotes the power imbalance at bus \(i\); Equation \eqref{equ:balance} enforce nodal power balance; Equation \eqref{equ:non-negative} and \eqref{equ:lineflow} pose hard constraints over mismatch nonnegativity and line flow directions respectively. Note such solved $b_i^g. f_{ij}$ can not fully represent optimal power dispatch solved from optimal power flow or economic dispatch. As time-varying nodal load, line parameter and exact power flow information are sensitive and missing, we emphasize that our framework finds a realistic but not real power dispatch values, which can serve as a proxy for Alberta grid operation pattern.

\begin{figure*}
    \centering
    \includegraphics[width=0.99\linewidth]{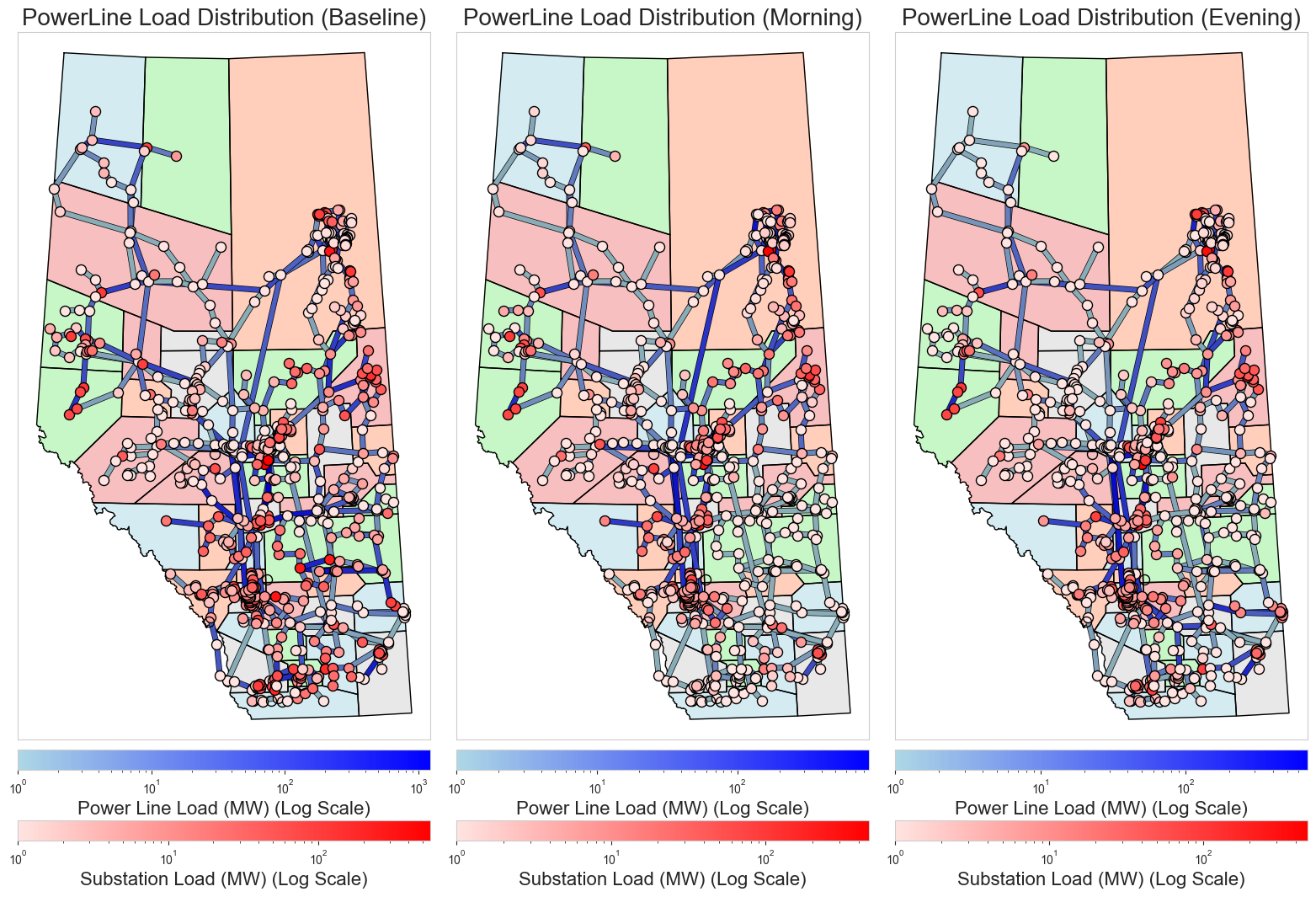}
    \caption{Bus and Powerline Load (a) Baseline; (b) Morning; (c) Evening. We overlay line and nodal load conditions using color gradients.}
    \label{fig:trend-within-day}
\end{figure*}

We find using our modeled topology and nodal demand, the LP module can output feasible solution, which successfully identify the global optimal solution for this objective with the error less than $1 \times 10^{-6}$. This indicates that we identify a powerline load allocation that exactly aligns with the bus load allocation, thereby producing a consistent allocation of $f_{i,j}$ for all power flows throughout the network.

\subsection{Graph-Based Grid Visualization}
In our study, we develop a static visualization of the power grid topology using Python. We use matplotlib for plotting, networkx for topology analysis, and cvxpy for linear programming to model load profiles. Power loads for buses and power lines are shown in different colors based on their intensity. While our current visualization produces static images to compare the grid at various time points, we plan to add dynamic, interactive features using tools like Plotly or Bokeh when continuous and multi-dimensional data become available.

\section{Periodic \& Trend Analysis}
Within our analytical framework, we employ two distinct modeling strategies to simulate power generation and distribution dynamics using AESO's load dataset:

\begin{enumerate}
    \item \textbf{Maximum Generator Capacity Mode:} This mode serves as the baseline for our analysis for analyzing grid conditions, where all generators are assumed to operate at their maximum efficiency and capacity.
    \item \textbf{Specific Time Point Data Mode:} This strategy models the generator operations based on real-time data fetched from a specific time point. 
\end{enumerate}

To quantify the impact of grid operating strategies under varying conditions on 2025-04-05, we analyze changes in power line directions when compared to the baseline. The visualization are organized together within Fig~\ref{fig:trend-within-day}, with gradient color indicating power level.

\begin{itemize}
    \item \textbf{Morning Dataset:} There are 94 (94/847) directional changes in power line flows observed in the morning compared to the baseline scenario under the Maximum Generator Capacity Mode.
    \item \textbf{Evening Dataset:} By evening of the same date, the number of directional changes from legacy mode increases to 91 (91/847), which is 71 directions different from the Morning data set, indicating slight variation in power distribution throughout the day.
\end{itemize}

Despite noting directional changes of approximately 10\% within a single day, the distribution visualization graphs remain largely consistent. This observation highlights that, although there are noticeable time variations in the distribution, they are limited in scope. These findings suggest that the majority of the power network maintains a steady configuration, while fluctuations come from demand changes and renewables variations.

\section{Conclusion and Future Works}
This study demonstrate that by leveraging linear programming and city population correlations along with public datasets on grid map and net demand, we can effectively reveal the overarching topology and power distribution within the Alberta grid. The approach, which incorporate Multi-source BFS-based methods for network exploration, enabled us to generate fine-grained visualizations despite limited information sources. These results highlight that with statistical and topological modeling techniques, incomplete datasets could be transformed into meaningful, actionable insights into the behavior and structure of complex power systems. Our approach is highly generalizable to other grids, and it provides a flexible tool for grid modeling and visualization.

In the future work, we plan to design approaches to identify congested lines/buses utilizing available public data. We would also like to investigate the interplay between congestion patterns, locational marginal prices (LMP), and locational carbon emission rates, aiming to provide strategies and analytical tools to improve grid efficiency and sustainability.

\bibliographystyle{IEEEtran}
\bibliography{bib}

\end{document}